\documentstyle[twoside]{article}

\catcode`\@=11
\long\def\@makefntext#1{
\protect\noindent \hbox to 3.2pt {\hskip-.9pt  
$^{{\eightrm\@thefnmark}}$\hfil}#1\hfill}		

\def\@makefnmark{\hbox to 0pt{$^{\@thefnmark}$\hss}}	
	
\def\ps@myheadings{\let\@mkboth\@gobbletwo
\def\@oddhead{\hbox{}
\rightmark\hfil\eightrm\thepage}   
\def\@oddfoot{}\def\@evenhead{\eightrm\thepage\hfil
\leftmark\hbox{}}\def\@evenfoot{}
\def\sectionmark##1{}\def\subsectionmark##1{}}

\oddsidemargin=\evensidemargin
\newcounter{sectionc}\newcounter{subsectionc}\newcounter{subsubsectionc}
\renewcommand{\section}[1] {\vspace{12pt}\addtocounter{sectionc}{1} 
\setcounter{subsectionc}{0}\setcounter{subsubsectionc}{0}\noindent 
	{\tenbf\thesectionc. #1}\par\vspace{5pt}}
\renewcommand{\subsection}[1] {\vspace{12pt}\addtocounter{subsectionc}{1} 
	\setcounter{subsubsectionc}{0}\noindent 
	{\bf\thesectionc.\thesubsectionc. {\kern1pt \bfit #1}}\par\vspace{5pt}}
\renewcommand{\subsubsection}[1] {\vspace{12pt}\addtocounter{subsubsectionc}{1}
	\noindent{\tenrm\thesectionc.\thesubsectionc.\thesubsubsectionc.
	{\kern1pt \tenit #1}}\par\vspace{5pt}}

\newcounter{appendixc}
\newcounter{subappendixc}[appendixc]
\newcounter{subsubappendixc}[subappendixc]
\renewcommand{\thesubappendixc}{\Alph{appendixc}.\arabic{subappendixc}}
\renewcommand{\thesubsubappendixc}
	{\Alph{appendixc}.\arabic{subappendixc}.\arabic{subsubappendixc}}

\renewcommand{\appendix}[1] {\vspace{12pt}
        \refstepcounter{appendixc}
        \setcounter{figure}{0}
        \setcounter{table}{0}
        \setcounter{lemma}{0}
        \setcounter{theorem}{0}
        \setcounter{corollary}{0}
        \setcounter{definition}{0}
        \setcounter{equation}{0}
        \renewcommand{\thefigure}{\Alph{appendixc}.\arabic{figure}}
        \renewcommand{\thetable}{\Alph{appendixc}.\arabic{table}}
        \renewcommand{\theappendixc}{\Alph{appendixc}}
        \renewcommand{\thelemma}{\Alph{appendixc}.\arabic{lemma}}
        \renewcommand{\thetheorem}{\Alph{appendixc}.\arabic{theorem}}
        \renewcommand{\thedefinition}{\Alph{appendixc}.\arabic{definition}}
        \renewcommand{\thecorollary}{\Alph{appendixc}.\arabic{corollary}}
        \renewcommand{\theequation}{\Alph{appendixc}.\arabic{equation}}
        \noindent{\tenbf Appendix \theappendixc #1}\par\vspace{5pt}}
\newcommand{\subappendix}[1] {\vspace{12pt}
        \refstepcounter{subappendixc}
        \noindent{\bf Appendix \thesubappendixc. {\kern1pt \bfit #1}}
	\par\vspace{5pt}}
\newcommand{\subsubappendix}[1] {\vspace{12pt}
        \refstepcounter{subsubappendixc}
        \noindent{\rm Appendix \thesubsubappendixc. {\kern1pt \tenit #1}}
	\par\vspace{5pt}}

\newcommand{\textlineskip}{\baselineskip=13pt}
\newcommand{\smalllineskip}{\baselineskip=10pt}

\def\eightcirc{
\begin{picture}(0,0)
\put(4.4,1.8){\circle{6.5}}
\end{picture}}
\def\eightcopyright{\eightcirc\kern2.7pt\hbox{\eightrm c}}

\def\abstracts#1#2#3{{
	\centering{\begin{minipage}{4.5in}\baselineskip=10pt\footnotesize
	\parindent=0pt #1\par 
	\parindent=15pt #2\par
	\parindent=15pt #3
	\end{minipage}}\par}} 

\newcommand{\bibit}{\nineit}

\renewenvironment{thebibliography}[1]
	{\frenchspacing
	 \ninerm\baselineskip=11pt
	 \begin{list}{\arabic{enumi}.}
	{\usecounter{enumi}\setlength{\parsep}{0pt}
	 \setlength{\leftmargin 12.7pt}{\rightmargin 0pt} 
	 \setlength{\itemsep}{0pt} \settowidth
	{\labelwidth}{#1.}\sloppy}}{\end{list}}

\newcounter{itemlistc}
\newcounter{romanlistc}
\newcounter{alphlistc}
\newcounter{arabiclistc}

\newcommand{\fcaption}[1]{
        \refstepcounter{figure}
        \setbox\@tempboxa = \hbox{\footnotesize Fig.~\thefigure. #1}
        \ifdim \wd\@tempboxa > 5in
           {\begin{center}
        \parbox{5in}{\footnotesize\smalllineskip Fig.~\thefigure. #1}
            \end{center}}
        \else
             {\begin{center}
             {\footnotesize Fig.~\thefigure. #1}
              \end{center}}
        \fi}

\newcommand{\tcaption}[1]{
        \refstepcounter{table}
        \setbox\@tempboxa = \hbox{\footnotesize Table~\thetable. #1}
        \ifdim \wd\@tempboxa > 5in
           {\begin{center}
        \parbox{5in}{\footnotesize\smalllineskip Table~\thetable. #1}
            \end{center}}
        \else
             {\begin{center}
             {\footnotesize Table~\thetable. #1}
              \end{center}}
        \fi}

\def\@citex[#1]#2{\if@filesw\immediate\write\@auxout
	{\string\citation{#2}}\fi
\def\@citea{}\@cite{\@for\@citeb:=#2\do
	{\@citea\def\@citea{,}\@ifundefined
	{b@\@citeb}{{\bf ?}\@warning
	{Citation `\@citeb' on page \thepage \space undefined}}
	{\csname b@\@citeb\endcsname}}}{#1}}

\newif\if@cghi
\def\cite{\@cghitrue\@ifnextchar [{\@tempswatrue
	\@citex}{\@tempswafalse\@citex[]}}
\def\citelow{\@cghifalse\@ifnextchar [{\@tempswatrue
	\@citex}{\@tempswafalse\@citex[]}}
\def\@cite#1#2{{$\null^{#1}$\if@tempswa\typeout
	{IJCGA warning: optional citation argument 
	ignored: `#2'} \fi}}

\def\pmb#1{\setbox0=\hbox{#1}
	\kern-.025em\copy0\kern-\wd0
	\kern.05em\copy0\kern-\wd0
	\kern-.025em\raise.0433em\box0}

\def\fnt#1#2{\footnotetext{\kern-.3em
	{$^{\mbox{\scriptsize #1}}$}{#2}}}

\def\fpage#1{\begingroup
\voffset=.3in
\thispagestyle{empty}\begin{table}[b]\centerline{\footnotesize #1}
	\end{table}\endgroup}

\font\tenrm=cmr10
\font\tenit=cmti10 
\font\tenbf=cmbx10
\font\bfit=cmbxti10 at 10pt
\font\ninerm=cmr9
\font\nineit=cmti9

\font\eightrm=cmr8

\def\qed{\hbox{${\vcenter{\vbox{			
   \hrule height 0.4pt\hbox{\vrule width 0.4pt height 6pt
   \kern5pt\vrule width 0.4pt}\hrule height 0.4pt}}}$}}

\begin{document}


\normalsize\textlineskip
\thispagestyle{empty}
\setcounter{page}{1}

\begin{flushright}
UB-HET-00-03\\
November 2000
\end{flushright}

\vspace*{0.1truein}

\fpage{1}
\centerline{\bf PROBING NEUTRAL GAUGE BOSON SELF-INTERACTIONS IN $ZZ$ }
\vspace*{0.035truein}
\centerline{\bf PRODUCTION AT THE TEVATRON}
\vspace*{0.37truein}
\centerline{\footnotesize U. BAUR}
\vspace*{0.015truein}
\centerline{\footnotesize\it Physics Department, State University
of New York at Buffalo}
\baselineskip=10pt
\centerline{\footnotesize\it Buffalo, NY 14260, USA}
\vspace*{10pt}
\centerline{\footnotesize D. RAINWATER}
\vspace*{0.015truein}
\centerline{\footnotesize\it Theory Group, Fermilab}
\baselineskip=10pt
\centerline{\footnotesize\it Batavia, IL 60510, USA}

\vspace*{0.21truein}
\abstracts{We present an analysis of $ZZ$ production at the upgraded
Fermilab Tevatron for general $ZZZ$ and $ZZ\gamma$ 
couplings. Achievable limits on these couplings 
are shown to be a significant improvement over the limits currently
obtained by LEP~II.}{}{}


\vspace*{1pt}\textlineskip	
\section{Introduction}	
\vspace*{-0.5pt}
\noindent
The Standard Model (SM) of electroweak interactions makes precise predictions 
for the couplings between gauge bosons due to the non-abelian gauge 
symmetry of 
$SU(2)_L\otimes U(1)_Y$. These self-interactions are described by the triple 
gauge boson (trilinear) $WWV$, $Z\gamma V$, and $ZZV$ ($V=\gamma,\,Z$) 
couplings and the quartic couplings. Vector boson pair production provides a 
sensitive ground for {\em direct tests} of the trilinear couplings. Deviations 
of the couplings from the expected values would indicate the presence of new 
physics beyond the SM.

To date the SM has passed this rigorous test with no observed deviations from 
the SM values. The $WWV$ and $Z\gamma V$ couplings have been measured 
with an accuracy of ${\cal O}(10\%)$ at LEP2 and the
Tevatron\cite{wudka}. The
$ZZV$ couplings, on the other hand, are only loosely constrained at the
moment through $ZZ$ production at LEP2~\cite{LEP2}. Due to low event
rates after branching ratios, or large backgrounds, $ZZ$ production was not
observed by the Tevatron experiments in Run~I. In Run~II of the Tevatron
which will begin in~2001, an integrated
luminosity of $2-15$~fb$^{-1}$ is envisioned,  and a sufficient number of $ZZ$ 
events should be available to commence a detailed investigation of the 
$ZZV$ couplings. In the following we summarize the results of a recent
detailed study\cite{BR} of the capabilities of future Tevatron
experiments to test the $ZZV$ vertices via $ZZ$ production.

\section{$\boldmath{ZZZ}$ and $\boldmath{ZZ\gamma}$ Anomalous Couplings}
\noindent
Two $ZZZ$ couplings, and two $ZZ\gamma$ couplings, are allowed by
electromagnetic gauge invariance and Lorentz invariance\cite{Wcoupling} 
for on-shell $Z$ bosons. In the massless fermion limit, the most general
form of the 
$Z^\alpha(q_1)\,Z^\beta(q_2)\, V^\mu(P)$ ($V=Z,\,\gamma$) vertex
function 
may be written as\cite{Wcoupling} 
\begin{equation}
g_{ZZV} \Gamma^{\alpha\beta\mu}_{ZZV} = 
e\, {P^2-M_V^2\over M_Z^2}\, \biggl[ if_4^V \left(P^\alpha
g^{\mu\beta}+P^\beta g^{\mu\alpha} \right) 
+if_5^V\epsilon^{\mu\alpha\beta\rho}\left(q_1-q_2\right)_\rho\biggr],
\label{eq:VVV}
\end{equation}
where $M_Z$ is the $Z$-boson mass and $e$ is the proton charge. 
The overall factor $(P^2-M_V^2)$ in
Eq.~(\ref{eq:VVV}) is a consequence of Bose symmetry for $ZZZ$
couplings, while it is due to electromagnetic gauge invariance for the
$ZZ\gamma$ couplings. All
couplings are $C$ odd; $CP$ invariance forbids $f^V_4$ and parity 
conservation requires that $f_5^V$ vanishes. In the SM, at tree level,
$f^V_4 = f^V_5 = 0$. 

$S$-matrix
unitarity restricts the $ZZV$ couplings uniquely to their SM values at
asymptotically high energies\cite{unitarity}. This requires that the 
couplings $f^V_i$ possess a momentum dependence which ensures
that the $f^V_i(\hat s)$ vanish for $\hat s\to\infty$. In order to avoid
unphysical results that would violate unitarity, the $\hat{s}$
dependence thus has to be taken into account. To parameterize the $\hat
s$ dependence of the form factor, 
we use a generalized dipole form factor, $f^V_i(\hat{s}) = 
f^V_{i0}/( 1 + \hat{s}/\Lambda_{FF}^{2})^{n}, (i = 4,5)$,
where $\Lambda_{FF}$ is the form factor scale which is related to the
scale of the new physics which is generating the anomalous $ZZV$
couplings. The values of the form factors at low energy, $f^V_{i0}$, and
the power of the form factor, $n$, are
constrained by partial wave unitarity of the inelastic $ZZ$ production
amplitude in fermion antifermion annihilation at arbitrary
center-of-mass energies\cite{BR}. 

\section{Signatures of Anomalous $ZZV$ Couplings}
\noindent
Our analysis examines the observable final state signatures, $ZZ\to 
\ell_1^+\ell_1^-\ell_2^+\ell_2^-$, 
$\ell^+\ell^-\nu\bar{\nu}$, $\ell^+\ell^- jj$
($\ell,\,\ell_1,\,\ell_2=e,\,\mu$) and $\bar\nu\nu jj$. The total
cross section for $p\bar p\to ZZ$ at $\sqrt{s}=2$~TeV, including NLO QCD 
corrections is approximately 1.5~pb. For an integrated 
luminosity of 2~fb$^{-1}$ one thus expects a few
$ZZ\to\ell_1^+\ell_1^-\ell_2^+\ell_2^-$ ($\ell_1,\,\ell_2=e,\,\mu$)
events, if realistic lepton $p_T$ and pseudo-rapidity
cuts are imposed. Larger event rates are expected for
$ZZ\to\ell^+\ell^-\bar\nu\nu$, $ZZ\to\ell^+\ell^- jj$ and
$ZZ\to\bar\nu\nu jj$. These channels, however, suffer from non-trivial
background contributions. 

The effects of anomalous 
$ZZV$ couplings are enhanced at large energies. A typical signal of
nonstandard $ZZZ$ and $ZZ\gamma$ couplings thus will be a broad increase 
in the $ZZ$ invariant mass distribution, the $Z$ transverse momentum
distribution and the $p_T$ distribution of the $Z$ decay leptons. 

The number of $ZZ\to\ell^+\ell^-\bar\nu\nu$ signal events is about a
factor~6 larger than the number of $ZZ\to 4$~leptons events. The two
most important backgrounds to $ZZ\to\ell^+\ell^-\bar\nu\nu$ production
are $t\bar t\to W^+W^-b\bar b$ and $W^+W^-\to\ell^+\nu\ell^-\bar\nu$
production. A jet veto almost completely eliminates the $t\bar t$
background. The $W^+W^-$ background exceeds the $ZZ$ signal cross 
section for $p_T(e^+e^-)<80$~GeV. The $p_T(e^+e^-)$ distribution of the 
$W^+W^-$ background, however, drops much faster than that of the $ZZ$
signal, and, thus, will only marginally affect the sensitivity to $ZZV$
couplings. 

The decay modes where one of the two $Z$ bosons decays hadronically have 
much larger branching fractions than the $ZZ\to 4$~leptons and the
$ZZ\to\ell^+\ell^-\bar\nu\nu$ channels, but also much higher
backgrounds. The main background sources are QCD $Z+2$~jet (``$Zjj$'')
production and $WZ$ production with $W\to jj$. 
In both cases the $jj$ invariant mass is constrained to be
near the $Z$ pole. Other potentially dangerous background sources are 
$t\bar{t}$ and $Wjj$
production. The $Wjj$ background contributes only to the 
$\bar\nu\nu jj$ final state. For $ZZ\to\ell^+\ell^- jj$, the $t\bar
t\to\ell^+\nu_\ell\ell^-\bar\nu_\ell \bar bb$ background can be reduced 
by requiring that the missing transverse momentum is
$p\llap/_T<20$~GeV. For $ZZ\to \bar\nu\nu jj$, suppression of the $t\bar
t\to\ell^+\nu_\ell\ell^-\bar\nu_\ell \bar bb$ and $W(\to\ell\nu)jj$ 
backgrounds is possible by
requiring that there are no central high-$p_T$ charged leptons 
present in the event. The $Zjj$ background is found to be, by far, the
largest background. Its size is uniformly about one 
order of magnitude 
larger than the SM $ZZ$ signal. It will therefore be very difficult to observe 
$ZZ$ production in the semi-hadronic channels, if the SM prediction 
is correct. 
However, for sufficiently large anomalous $ZZV$ couplings, the $ZZ$ cross 
section exceeds the background at large transverse momenta. The
semi-hadronic channels therefore may still be useful 
in obtaining limits on the $ZZV$ couplings at the Tevatron. 

\section{Sensitivity Limits}
\noindent
In order to derive sensitivity limits for anomalous $ZZV$ couplings
which one can hope to achieve in Run~II, we use the $p_T(\ell^+\ell^-)$
distribution for $ZZ\to 4$~leptons, $ZZ\to\ell^+\ell^-\bar\nu\nu$ and
$ZZ\to\ell^+\ell^- jj$. For the $ZZ\to\bar\nu\nu jj$ channel we use the
$p_T(jj)$ distribution. Other distributions, such as the $ZZ$ invariant
mass distribution (useful only for $ZZ\to 4$~leptons), or the maximum or 
minimum transverse momenta of the charged leptons or jets, yield similar
results. In deriving our sensitivity limits, we combine channels with
electrons and muons in the final state. We calculate 95\% confidence 
level (CL) limits performing a $\chi^2$ test, allowing for 
a normalization uncertainty of 30\% of the SM cross section. The most
stringent bounds at Tevatron energies are obtained from
$ZZ\to\ell^+\ell^-\bar\nu\nu$ and $ZZ\to\bar\nu\nu jj$. For $\int\!{\cal
L}dt=2~{\rm fb}^{-1}$, $n=3$ and $\Lambda_{FF}=750$~GeV one finds:
\begin{eqnarray}
|f^Z_{40}|<0.159 &\qquad & -0.184< f^Z_{50}<0.162 \\
|f^\gamma_{40}|<0.163 &\qquad & -0.179< f^\gamma_{50}<0.170.
\end{eqnarray}
These bounds improve the present limits from LEP~II\cite{LEP2} by a
factor~3 to~6. The limits from 
the $ZZ\to\ell^+\ell^- jj$ and $ZZ\to 4$~leptons channels are about a 
factor~1.5 and~2 weaker than those from $ZZ\to\ell^+\ell^-\bar\nu\nu$
and $ZZ\to\bar\nu\nu jj$. Bounds roughly scale with $(\int\!{\cal
L}dt)^{1/4}$. At the LHC one will be able to probe $ZZV$ couplings of
${\cal O}(10^{-3})$ in magnitude\cite{BR}.

\end{document}